\documentclass[modern]{aastex631}
\usepackage{amsmath,amssymb,CJK,todonotes,pifont}

\shorttitle{P/2021 HS (PANSTARRS) and the Low-Activity Comets}
\shortauthors{Ye et al.}

\begin{document}
\begin{CJK*}{UTF8}{gbsn}

\title{Comet P/2021 HS (PANSTARRS) and the Challenge of Detecting Low-Activity Comets}

\correspondingauthor{Quanzhi Ye}
\email{qye@umd.edu}

\author[0000-0002-4838-7676]{Quanzhi Ye (叶泉志)}
\affiliation{Department of Astronomy, University of Maryland, College Park, MD 20742, USA}
\affiliation{Center for Space Physics, Boston University, 725 Commonwealth Ave, Boston, MA 02215, USA}

\author[0000-0002-6702-7676]{Michael S. P. Kelley}
\affiliation{Department of Astronomy, University of Maryland, College Park, MD 20742, USA}

\author[0000-0001-9542-0953]{James M. Bauer}
\affiliation{Department of Astronomy, University of Maryland, College Park, MD 20742, USA}

\author[0000-0002-4767-9861]{Tony L. Farnham}
\affiliation{Department of Astronomy, University of Maryland, College Park, MD 20742, USA}

\author[0000-0002-2668-7248]{Dennis Bodewits} 
\affiliation{Department of Physics, Edmund C. Leach Science Center, Auburn University, Auburn AL 36832, USA}

\author{Luca Buzzi}
\affiliation{Schiaparelli Astronomical Observatory, Varese, Italy}



\author[0000-0002-0439-9341]{Robert Weryk}
\affiliation{Department of Physics and Astronomy, The University of Western Ontario, London, ON N6A 3K7, Canada}



\author[0000-0002-8532-9395]{Frank J. Masci}
\affiliation{IPAC, California Institute of Technology, 1200 E. California Blvd, Pasadena, CA 91125, USA}

\author[0000-0002-7226-0659]{Michael S. Medford}
\affiliation{Department of Astronomy, University of California, Berkeley, Berkeley, CA 94720, USA}
\affiliation{Lawrence Berkeley National Laboratory, 1 Cyclotron Rd., Berkeley, CA 94720, USA}

\author[0000-0002-0387-370X]{Reed Riddle}
\affiliation{Caltech Optical Observatories, California Institute of Technology, Pasadena, CA 91125, USA}

\author[0000-0002-9998-6732]{Avery Wold}
\affiliation{IPAC, California Institute of Technology, 1200 E. California Blvd, Pasadena, CA 91125, USA}





\begin{abstract}

Jupiter-family comet (JFC) P/2021 HS (PANSTARRS) only exhibits a coma within a few weeks of its perihelion passage at 0.8~au, which is atypical for a comet. Here we present an investigation into the underlying cause using serendipitous survey detections as well as targeted observations. We find that the detection of the activity is caused by an extremely faint coma being enhanced by forward scattering effect due to the comet reaching a phase angle of $\sim140^\circ$. The coma morphology is consistent with sustained, sublimation-driven activity produced by a small active area, $\sim700~\mathrm{m^2}$, one of the smallest values ever measured on a comet. The phase function of the nucleus shows a phase coefficient of $0.035\pm0.002~\mathrm{mag/deg}$, implying an absolute magnitude of $H=18.31\pm0.04$ and a phase slope of $G=-0.13$, with color consistent with typical JFC nuclei. Thermal observations suggest a nucleus diameter of 0.6--1.1~km, implying an optical albedo of 0.04--0.23 which is higher than typical cometary nuclei. An unsuccessful search for dust trail and meteor activity confirms minimal dust deposit along the orbit, totaling $\lesssim10^8$~kg. As P/2021 HS is dynamically unstable, similar to typical JFCs, we speculate that it has an origin in the trans-Neptunian region, and that its extreme depletion of volatiles is caused by a large number of previous passages to the inner Solar System. The dramatic discovery of the cometary nature of P/2021 HS highlights the challenges of detecting comets with extremely low activity levels. Observations at high phase angle where forward scattering is pronounced will help identify such comets.

\end{abstract}



\section{Introduction}

Comets are unaccreted ice-rich planetesimals originated in the outer Solar System. They formed at different locations within the protoplanetary disk and have undergone various evolutionary paths, thus exhibiting a wide range of observational properties. For instance, some comets produce significantly more dust and/or gas than the average, while some behave the opposite. Such a diversity can be explained by the degree of depletion of volatiles and near-surface dust due to aging as well as the abundance of volatiles in the birth environment, and thus understanding of the cause of hyper- or hypo-activity of comets -- i.e., the storage and release of volatiles -- furnishes our knowledge of the origin and evolution of comets. However, such efforts are often hindered by the chaotic dynamical evolution of comets and the limited interpretation permitted by Earth-based or even spacecraft observations \citep[e.g.,][]{Choukroun2020}. The orbits of most short-period comets can only be reliably traced back for $10^3$--$10^4$~yrs \citep{Tancredi2014}, a negligible fraction of their lifetime. The behavior of comets can also change dramatically over a few orbital revolutions \citep[e.g. 240P/NEAT, 252P/LINEAR,][]{Kelley2019, Li2017}, making it difficult to assess the comet's volatile content based on observations of limited orbits.

Comet P/2021 HS (PANSTARRS) was discovered by the Panoramic Survey Telescope and Rapid Response System (Pan-STARRS) survey on 2021 April 16, initially as an asteroid \citep{MPC2021}. It has perihelion distance $q=0.80$~au and Jupiter Tisserand parameter $T_\mathrm{J}=2.27$, hence simultaneously fitting into the definitions of a near-Earth comet and a Jupiter-family comet (JFC). With a Minimum Orbit Intersection Distance of 0.04~au with the Earth, as well as an absolute magnitude of 20.6, the comet is also classified as a potentially hazardous object (PHO). P/2021 HS was found to exhibit a coma in images taken in 2021 July and August, a few weeks from its perihelion passage \citep{CBET5043}. Here we present our analyses of data collected by a collection of time-domain surveys and dedicated observing campaigns, with the goal of understanding the current state, activity mechanism, and nature of P/2021 HS.

\section{Observations}

We assembled an array of data of P/2021 HS from serendipitous survey detections as well as targeted observations, as described below and summarized in Table~\ref{tbl:obs}.

\subsection{Dark Energy Camera (DECam)}

DECam is a wide-field camera mounted on the 4-m Victor M. Blanco telescope at the Cerro Tololo Inter-American Observatory. The camera has a hexagon-shaped, $3~\mathrm{deg}^2$-wide field-of-view and a pixel scale of $0.26''$.

We searched the DECam image catalog using the Canadian Astronomy Data Centre's Solar System Object Image Search \citep{Gwyn2012} and found 7 images taken from 2021 January 31 to June 1 that potentially contain P/2021 HS, all of which were taken in the course of the DECam eROSITA Survey (DeROSITAS)\footnote{\url{http://astro.userena.cl/derositas/}}. Upon closer examination, we found that only 3 out of the 7 images were usable for our purpose; for the rest of the images, the comet was either in the chip gaps or fell off the edge. Among these 3 usable images, the comet itself was only detected in a $150$~s $Y$-band image taken on 2021 March 27, and was not visible on the two $u$-band images taken on 2021 January 23 and March 21. By using the Guide Star Catalog Version 2.4.2 \citep{Lasker2021}, we established an upper brightness limit of $u>22.5$~mag for the comet at the two non-detection epochs.

\subsection{Near-Earth Object Wide-field Infrared Survey Explorer (NEOWISE)}

{\it NEOWISE} is repurposed from the Wide-field Infrared Survey Explorer ({\it WISE}) mission with the primary goal of discovering and characterizing Near-Earth Objects (NEOs). It operates in two infrared channels with central wavelengths of 3.4~\micron{} (W1) and 4.6~\micron{} (W2) with a fixed exposure time of 7.7~s. {\it NEOWISE} exclusively operates in survey mode, and thus detections of Solar System objects are untargeted. To facilitate moving object detection, the telescope adopts a survey cadence that repeatedly visits the same part of the sky over a period of $\sim36$~hr, and thus detections of a given object are grouped in date. For our case, P/2021 HS was serendipitously detected on 2021 January 13/14, May 28/29, and September 18/19.

The co-added images of each epoch were created from single images using the routine described in \citet{Masci2009}. We then performed aperture photometry on the comet using an aperture with $11''$ (equivalent to 21000, 5000 and 3200~km at the comet on the three epochs) in radius. The background was computed using annuli apertures centered on the comet with inner and outer radii of $30''$ and $50''$. The flux was calibrated following the procedure described in \citet{Bauer2015, Bauer2017}.

\subsection{Panoramic Survey Telescope \& Rapid Response System (Pan-STARRS)}

The Pan-STARRS survey consists of two 1.8-m telescopes located in Haleakala, Hawaii \citep{Chambers2016}. Each telescope is equipped with a $7~\mathrm{deg^2}$ camera and has a $0.25''$ pixel scale, complete with multiple broad-band photometric filters. Typical exposure times are $45$~s, allowing the system to reach $V\sim23$ mag.

We searched the archive back to 2021 January 1. To exclude poor images, we initially required the image full-width-half-maximum (FWHM) to be $<2''$. Images of P/2021 HS were found in the nights of 2021 April 16, 18, 22; May 14; and November 11. The comet was star-like in all these images, with a FWHM within $\sim1\sigma$ compared to background stars. In addition, we found three images from 2021 July 18 (one in $i$) and 20 (two in $r$ and $i$), after which time cometary activity had been reported, but the comet was near the detection limit owing the poor sky condition during the observation and no conclusion can be made about the activity.

\subsection{Transiting Exoplanet Survey Satellite (TESS)}

{\it TESS} is a space telescope with a primary goal to look for transiting exoplanets \citep{Ricker2015}. It carries four wide-field cameras, each covers a $24~\mathrm{deg^2}$ field-of-view at a pixel scale of $21''$. The large field-of-view and the large pixel sizes has proven useful for the search of low surface brightness structures such as faint comet trails \citep[e.g., the trail of 46P/Wirtanen,][]{Farnham2019}. P/2021 HS transited {\it TESS} field Sector 37 during 2021 April 2--28. We found a total of 3,446 images that contained P/2021 HS, each with an exposure time of 10 minutes, hence providing a total integration time of 574~hours. We stacked all these images on the comet to generate  final co-added images for further analysis.  We approached the stacking with two methods. The first of which processes the data to remove the static (non-time varying) sky. This technique tends to affect large extended features, such as comae, tails, and trails. The second approach aligns the data in the rest frame of the comet, rotating the data to fix the projected heliocentric velocity vector along the $-x$-axis.  The data are median combined. To remove star trails, the median-combined images are convolved with a 3~pix$\times$41~pix kernel oriented along the trailing direction. The project orbit is masked (6-pix width) before the convolution, and the result is subtracted from the median-combined image.  This approach has more noise and artifacts from the stars and filtering, but should better preserve trail photometry. A comparison between the two methods for an active comet with a dust trail was done by \citet{Farnham2019}.

\subsection{The Zwicky Transient Facility (ZTF)}

ZTF is a time-domain, $gri$-band optical survey conducted using the 1.2~m Oschin Schmidt telescope at Palomar Observatory \citep{Bellm2019, Graham2019}. Each image covers $47~\mathrm{deg^2}$ with a pixel scale of $1.0''$. A typical survey image has an exposure time of 30~s and a depth of $r\sim21$~mag.

We searched for images of P/2021 HS in the ZTF Data Release 10 and proprietary partnership data using {\tt ZChecker} \citep{Kelley2019b}. To exclude images in which the comet would be too faint or the position of the comet would be too uncertain, we focus on images taken after 2021 January 1 (i.e. 4 months before the first reported detection of the comet) with FWHMs $<3''$. Data Release 10 includes public data up to 2022 January 5. We found a total of 55 images of the comet, covering a period between 2021 May 11 and November 13. The photometry of the comet was then measured using a uniform $4''$-radius (varies between 2100 and 800~km at the comet), calibrated to the Pan-STARRS 1 (PS1) Data Release 1 (DR1) catalog \citep{Chambers2016} using ZTF's own pipeline \citep{Masci2019}. The reason behind using a fixed-angular aperture instead of a fix-distance aperture is that the latter would vary by $2.5\times$ throughout the period of search, with the times that the aperture size is largest coincide with the times that the images are most affected by twilight, making it difficult to derive high-quality photometry. {\tt ZChecker} also provides median-combined nightly stacked images that can be used for a deeper search of coma as well as characterizing coma morphology.

\subsection{Lowell Discovery Telescope (LDT)}

We observed P/2021 HS using the 4.3~m LDT on UT 2021 November 9, and December 6. Images were obtained using LDT's Large Monolithic Imager \citep[LMI;][]{Massey2013}. LMI has a field-of-view of $12\farcm3\times12\farcm3$ and a pixel scale of 0\farcs36 after on-chip $3\times3$ binning. The main purpose of the LDT imaging was to search for comet activity, hence all images were obtained through the \textit{VR} filter in order to maximize sensitivity. Images were bias-subtracted, flat-field corrected, and were then median-combined into stacked images. These images were photometrically calibrated to the PS1 DR1 catalog \citep{Chambers2016} with the bandpass approximated as $r'$ using {\tt PHOTOMETRYPIPELINE} \citep{Mommert2017}. The photometry of the comet was then measured using a uniform $4''$-radius consistent with the ZTF photometry discussed above.

\subsection{iTelescope -- Siding Spring}

iTelescope is a network of telescopes located at five locations around the world\footnote{\url{https://www.itelescope.net}.}. Activity of P/2021 HS was first detected using a 0.5-m astrograph at iTelescope -- Siding Spring site on 2021 July 4 and 6 \citep{CBET5043}. These images were taken through a Luminance ``$L$'' filter (an ultra-broadband filter that roughly covers $g$ and $r$) and have a pixel scale of 1\farcs6 after binned by 2. Images were then bias-subtracted, flat-field corrected and median-combined into two nightly images.

\subsection{Skygems Observatories -- Namibia}

The Skygems Observatories houses several telescopes in Spain and Namibia\footnote{\url{https://skygems-observatories.com}}. Observations of P/2021 HS were made using a 0.5-m astrograph at its Namibia location on 2021 July 9--10. Similar to the iTelescope observations, these images were taken through an $L$ filter. The pixel scale is 1\farcs1 after a $2\times2$ binning. The data were bias-subtracted, flat-field corrected and median-combined into two nightly images.

\begin{longrotatetable}
\begin{deluxetable*}{lccccccccl}
\tablecaption{Summary of observations used in this study.}
\label{tbl:obs}
\tablecolumns{8}
\tablehead{
\colhead{Date/Time (UT)} & \colhead{$r_\mathrm{h}$ (au)} & \colhead{$\varDelta$ (au)} & \colhead{$\alpha$} & $T-T_\mathrm{p}$ (d) & \colhead{Telescope} & \colhead{Filter} & \colhead{Exposure} & \colhead{Comment}
}

\startdata
2021 Jan 13--14 & 2.80 & 2.62 & $21^\circ$ & -206 to -205 & {\it NEOWISE} & W1, W2 & $14\times7.7$~s, $14\times7.7$~s & Nondetection \\
2021 Jan 23 08:17 & 2.71 & 2.39 & $21.0^\circ$ & -195.1 & DECam & $u$ & 60~s & Nondetection \\
2021 Mar 21 07:44 & 2.14 & 1.24 & $14.7^\circ$ & -138.1 & DECam & $u$ & 60~s & Nondetection \\
2021 Mar 27 02:57 & 2.08 & 1.15 & $13.1^\circ$ & -132.3 & DECam & $Y$ & 150~s & \\
2021 Apr 2--28 & 2.0--1.6 & 1.1--0.8 & $13^\circ$--$15^\circ$ & -127 to -101 & {\it TESS} & {\it TESS} & $3446\times600$~s & \\
2021 Apr 16 8:49--9:41 & 1.86 & 0.89 & $12.4^\circ$ & -112.0 & Pan-STARRS & $w$ & $4\times45$~s & \\
2021 Apr 18 7:00--7:45 & 1.84 & 0.87 & $13.1^\circ$ & -110.1 & Pan-STARRS & $w$ & $4\times45$~s & \\
2021 Apr 22 7:01--8:13 & 1.79 & 0.84 & $14.9^\circ$ & -106.1 & Pan-STARRS & $i$ & $6\times45$~s &  \\
2021 May 11 04:35 & 1.58 & 0.71 & $28.5^\circ$ & -87.2 & ZTF & $r$ & $1\times30$~s & \\
2021 May 14 06:22--07:10 & 1.54 & 0.70 & $31.1^\circ$ & -84.1 & Pan-STARRS & $w$ & $4\times45$~s & \\
2021 May 28--29 & 1.37 & 0.63 & $44^\circ$ & -70 to -69 & {\it NEOWISE} & W1, W2 & $11\times7.7$~s, $11\times7.7$~s & \\ 
2021 May 30 04:30--04:34 & 1.36 & 0.63 & $45.1^\circ$ & -68.2 & ZTF & $r$, $i$ & 30~s, 30~s & \\
2021 Jun 6 04:26 & 1.28 & 0.61 & $51.5^\circ$ & -61.2 & ZTF & $g$ & 30~s, 30~s & \\
2021 Jul 4 19:36--19:44 & 0.97 & 0.47 & $82.0^\circ$ & -32.6 & iTelescope & $L$ & $7\times60$~s & Active \\
2021 Jul 6 19:18--19:29 & 0.95 & 0.46 & $84.8^\circ$ & -30.6 & iTelescope & $L$ & $10\times60$~s & Active \\
2021 Jul 9 18:02--18:48 & 0.93 & 0.44 & $89.0^\circ$ &  -27.6& SkyGems & $L$ & $30\times60$~s & Active \\
2021 Jul 10 17:30--19:15 & 0.92 & 0.43 & $90.5^\circ$ & -26.7 & SkyGems & $L$ & $30\times60$~s & Active \\
2021 Aug 1 03:51--03:58 & 0.80 & 0.29 & $129.8^\circ$ & -5.2 & ZTF & $r$ & $3\times30$~s & Active \\
2021 Aug 2 03:50--04:00 & 0.80 & 0.29 & $131.5^\circ$ & -4.2 & ZTF & $r$ & $3\times30$~s & Active \\
2021 Aug 4 03:48--04:01 & 0.80 & 0.28 & $134.6^\circ$ & -2.2 & ZTF & $r$ & $2\times30$~s & Active \\
2021 Sep 5 10:20--11:43 & 0.95 & 0.34 & $90.5^\circ$ & +30.1 & ZTF & $r$, $g$ & 30~s, 30~s & \\
2021 Sep 8 10:56--11:46 & 0.97 & 0.35 & $85.2^\circ$ & +33.1 & ZTF & $g$, $r$ & $2\times30$~s, 30~s & \\
2021 Sep 10 11:11--11:49 & 0.99 & 0.36 & $81.8^\circ$ & +35.1 & ZTF & $g$, $r$ & 30~s, 30~s & \\
2021 Sep 12 11:22--11:50 & 1.01 & 0.37 & $78.5^\circ$ & +37.1 & ZTF & $r$, $g$ & 30~s, 30~s & \\
2021 Sep 15 10:59--11:45 & 1.04 & 0.38 & $73.8^\circ$ & +40.1 & ZTF & $g$, $r$ & 30~s, 30~s & \\
2021 Sep 16 11:14 & 1.05 & 0.39 & $72.3^\circ$ & +41.1 & ZTF & $r$ & 30~s & \\
2021 Sep 17 10:56 & 1.06 & 0.39 & $70.8^\circ$ & +42.1 & ZTF & $g$ & 30~s & \\
2021 Sep 18--19 & 1.07 & 0.40 & $69^\circ$ & +43 to +44 &{\it NEOWISE} & W1, W2 & $8\times7.7$~s, $8\times7.7$~s & \\ 
2021 Sep 19 11:44 & 1.09 & 0.40 & $67.8^\circ$ & +44.1 & ZTF & $g$ & 30~s & \\
2021 Sep 21 10:17 & 1.11 & 0.41 & $65.1^\circ$ & +46.0 & ZTF & $i$ & 30~s & \\
2021 Sep 22 10:24 & 1.12 & 0.41 & $63.7^\circ$ & +47.0 & ZTF & $r$ & 30~s & \\
2021 Sep 30 09:24 & 1.20 & 0.44 & $52.8^\circ$ & +55.0 & ZTF & $r$ & 30~s & \\
2021 Oct 2 09:25--10:21 & 1.23 & 0.45 & $50.1^\circ$ & +57.0 & ZTF & $g$, $r$ & $2\times30$~s, $2\times30$~s & \\
2021 Oct 4 06:25--06:28 & 1.23 & 0.45 & $50.1^\circ$ & +58.9 & LDT & \textit{VR} & $3\times60$~s & \\
2021 Oct 9 09:00--09:47 & 1.31 & 0.48 & $41.2^\circ$ & +64.0 & ZTF & $g$, $r$ & 30~s, 30~s & \\
2021 Oct 15 10:55--11:56 & 1.38 & 0.51 & $33.7^\circ$ & +70.1 & ZTF & $g$, $r$ & 30~s, 30~s & \\
2021 Oct 17 08:49--11:15 & 1.40 & 0.52 & $31.4^\circ$ & +72.0 & ZTF & $i$, $g$, $r$ & 30~s, 30~s, 30~s & \\
2021 Oct 30 06:14--08:18 & 1.55 & 0.60 & $17.5^\circ$ & +84.9 & ZTF & $g$, $r$ & 30~s, 30~s & \\
2021 Nov 3 06:53--07:25 & 1.60 & 0.63 & $14.1^\circ$ & +88.9 & ZTF & $r$, $g$ & $2\times30$~s, $2\times30$~s & \\
2021 Nov 4 06:58--07:00 & 1.61 & 0.64 & $13.3^\circ$ & +89.9 & ZTF & $i$ & $2\times30$~s & \\
2021 Nov 5 06:23 & 1.62 & 0.65 & $12.6^\circ$ & +90.9 & ZTF & $r$ & 30~s & \\
2021 Nov 7 07:20--07:50 & 1.64 & 0.67 & $11.2^\circ$ & +92.9 & ZTF & $r$, $g$ & 30~s, 30~s & \\
2021 Nov 9 05:43--06:51 & 1.66 & 0.69 & $10.2^\circ$ & +94.9 & ZTF & $g$, $r$ & 30~s, 30~s & \\
2021 Nov 11 06:35--06:52 & 1.69 & 0.71 & $9.3^\circ$ & +96.9 & Pan-STARRS & $w$ & $2\times45$~s & \\
2021 Nov 12 08:17 & 1.70 & 0.73 & $9.1^\circ$ & +98.0 & ZTF & $r$ & 30~s & \\
2021 Nov 13 05:08 & 1.71 & 0.73 & $8.9^\circ$ & +98.8 & ZTF & $i$ & 30~s & \\
2021 Dec 6 06:01--06:05 & 1.97 & 1.06 & $15.5^\circ$ & +121.9 & LDT & \textit{VR} & $2\times180$~s & \\
\enddata
\tablecomments{$r_\mathrm{h}$, $\varDelta$, and $\alpha$ are heliocentric distance, geocentric distance, and phase angle of the comet, respectively.}
\end{deluxetable*}
\end{longrotatetable}

\section{Results}

\subsection{Nucleus and Gas Production}
\label{sec:neowise}

\begin{figure*}
\begin{center}
\includegraphics[scale=0.5,angle=0]{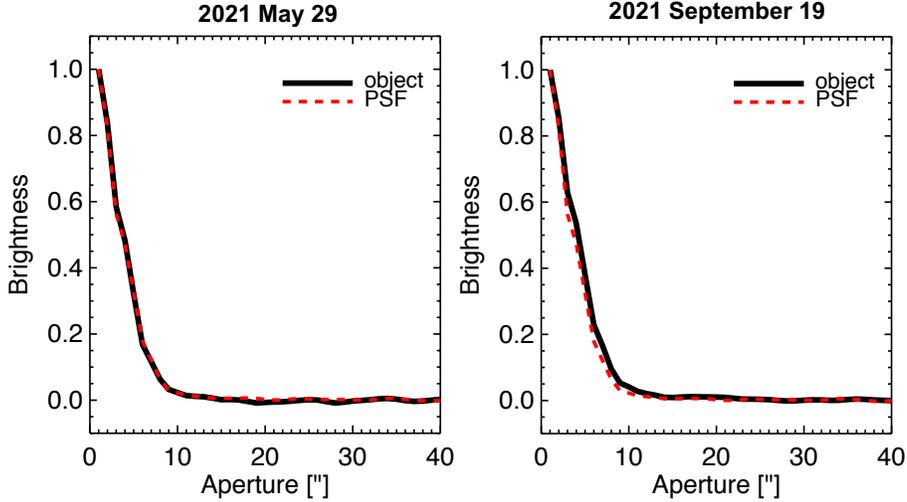}
\caption{NEOWISE W2 Surface brightness profiles of P/2021 HS on 2021 May 29 (left panel) and September 19 (right panel) as measured by {\it NEOWISE}. Neither show significant coma was detected. The error bars are too small to show in the figure.\label{fig:neowise-sb}}
\end{center}
\end{figure*}

{\it NEOWISE}'s infrared capabilities enable us to probe either the CO+CO$_2$ gas coma or the bare nucleus through its thermal radiation (if the gas emission is negligible). For the three dates that P/2021 HS were serendipitously observed, the comet was only visible on the last two dates, 2021 May 29 and September 19, without a visible coma or a tail/trail. Figure~\ref{fig:neowise-sb} shows the surface brightness profiles of P/2021 HS for these two dates in which the comet was detected. On both dates, the surface brightness profile of the comet closely follows the point-spread function, indicating that no substantial coma was present.

To derive the size of the nucleus, we employ the Near-Earth Asteroid Thermal Model (NEATM) developed by \citet{Harris1998}, a model that simplistically assumes a spherical nucleus with negligible thermal contribution from the nightside. Such simplifications are necessary as the rotational and thermal properties of the nucleus are virtually unknown; in addition, we will soon show that the exclusion of nighttime emission does not significantly impact our final result. With the model, we connect the nucleus size and beaming parameter $\eta$ (which generalize a number of thermal-related effects) to the measured thermal flux. From measurements at three different epochs (orbital positions), we determine $\eta=1.0\pm0.2$, in agreement with previously measured beaming parameters for comets \citep{Fernandez2013, Bauer2017}. The corresponding nucleus diameter is therefore 0.6--1.1~km. We note that such uncertainty range is significantly larger than the systematic bias of NEATM determined by \citet[][$+11\pm8\%$]{Wolters2009}, thus justifying our decision to ignore the nightside emission. Although the impact of the flux variation caused by the rotation of a (likely) irregular-shape body is unknown, we will see in the later section that the scatter of the optical light-curve is $\lesssim0.5$~mag, and therefore we conclude that such flux variation is unlikely to dominate the uncertainty of the calculated size.

Although no detection was found for the 2021 January 13--14 stack of {\it NEOWISE} images (see Table~\ref{tbl:obs}), upper limits of CO and CO$_2$ production were obtained from the 4.6 $\mu$m signal using the same production models described in \cite{Bauer2021}. The 3-$\sigma$ upper limits were found to be $2.1\times10^{25}$ molecules~s$^{-1}$ for CO$_2$, and $2.2\times10^{26}$ molecules~s$^{-1}$ for CO, no larger than the lowest values measured on previous of comets {\it NEOWISE} detections \citep{Bauer2015}, though the detection limit is above the average of JFC detections recently measured by \citet{Pinto2022}.

\subsection{Lightcurve}
\label{sec:lc}

Since P/2021 HS is officially classified as a comet, we first investigate its photometric behavior using the $Af\rho$ quantity \citep{Ahearn1984}. The $Af\rho$ quantity is a product of dust albedo $A$, aperture filling factor $f$, and the linear scale of the aperture $\rho$. Assuming that the observed comet flux is dominated by the flux from the dust coma, the $Af\rho$ quantity provides an approximation of the comet's dust production rate. Mathematically, it is defined as

\begin{equation}
    A(\alpha)f\rho = \frac{4 r_\mathrm{h}^2 \varDelta^2}{\rho} \frac{F_\mathrm{C}}{F_\odot}
\end{equation}

\noindent where $A(\alpha)f\rho$ means that the measurement is made for a phase angle of $\alpha$, $r_\mathrm{h}$ and $\varDelta$ is the heliocentric and geocentric distances of the comet, $F_\mathrm{C}$ is the spectral flux density of the coma, and $F_\odot$ is the spectral flux density of the Sun at 1~au. A phase angle and distance-independent quantity $A(0^\circ)f\rho$ can be computed by

\begin{equation}
    A(0^\circ)f\rho = \frac{A(\alpha)f\rho}{\Phi(\alpha)}
\end{equation}

\noindent where $\Phi(\alpha)$ is the phase function of the dust of which we model using the Schleicher--Marcus function \citep[][also called Halley--Marcus function in some literature]{Schleicher1998,Marcus2007}.

\begin{figure*}
\begin{center}
\includegraphics[scale=1,angle=0]{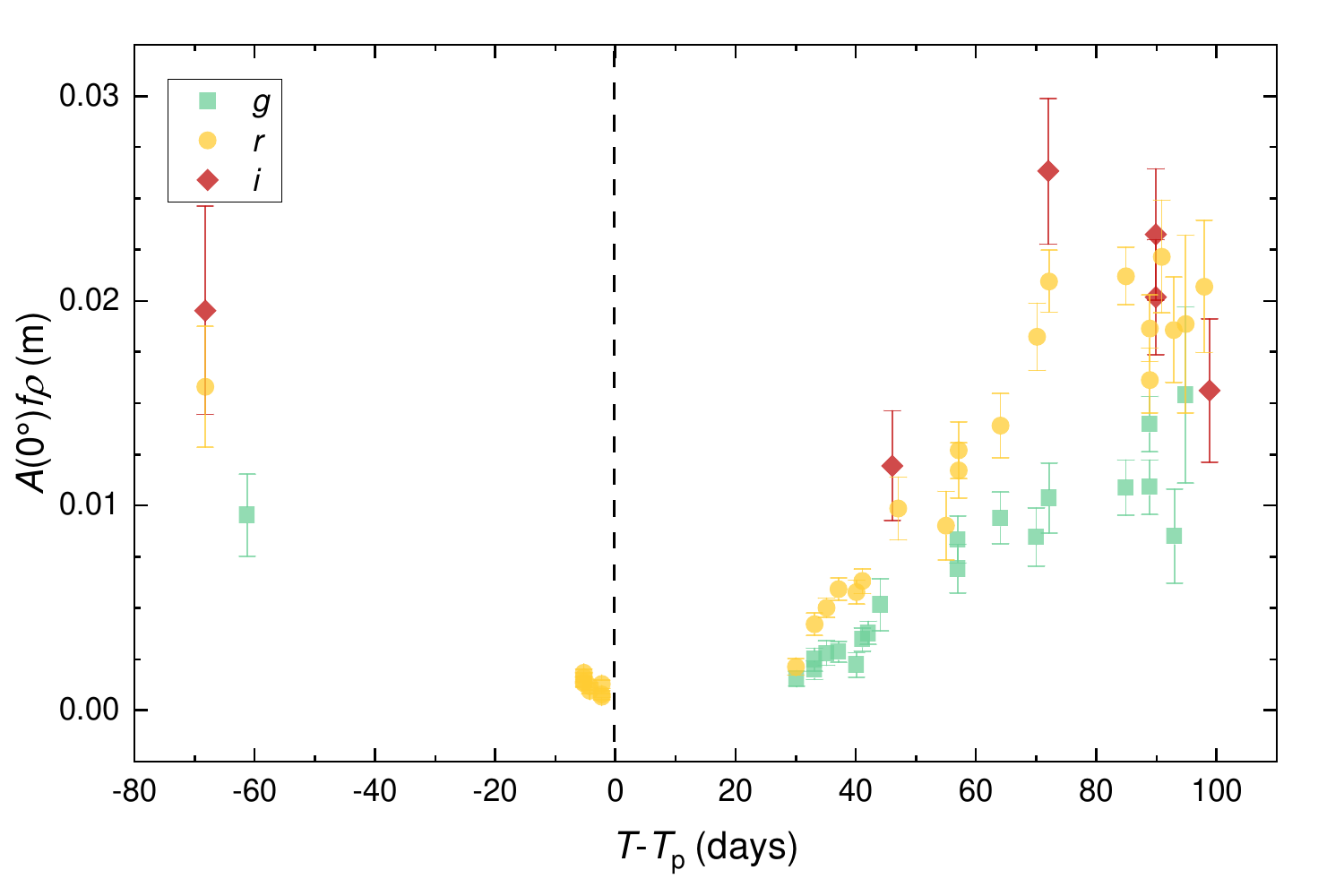}
\caption{Dust production metric $A(0^\circ)f\rho$ of P/2021 HS as a function of time, expressed as days to perihelion $T-T_\mathrm{p}$, as derived from $g$, $r$ and $i$-band ZTF photometric measurements. \label{fig:afrho}}
\end{center}
\end{figure*}

For consistency, we only used ZTF photometry as ZTF data is calibrated and provided good coverage of the comet's 2021 apparition. The computed $A(0^\circ)f\rho$ as a function of time is shown in Figure~\ref{fig:afrho}. Curiously, this figure shows a behavior in exactly the opposite of typical comets: instead of getting more active (i.e. intrinsically brighter) as the comet approaches the perihelion, P/2021 HS is getting less active approaching perihelion, reaching a minimum $A(0^\circ)f\rho$ of $\sim0.001$~m, only to be bouncing back after perihelion. The activity level of a comet is typically inversely related to heliocentric distance of the comet, since comet activity is usually driven by the sublimation of volatile material which is proportional to the solar flux received at the comet; no known physical mechanism can explain the exactly opposite behavior.

It then came to our attention that P/2021 HS reached maximum phase angle of $\alpha=139^\circ$ as seen from Earth only three days after perihelion. Such unusual geometry can potentially lead to significant brightness enhancement if the coma is dominated by dust, due to forward scattering effect. A complication can also arise if the dust is not a significant component, since the gas coma does not experience forward scattering effect.

\begin{figure*}
\begin{center}
\includegraphics[scale=1,angle=0]{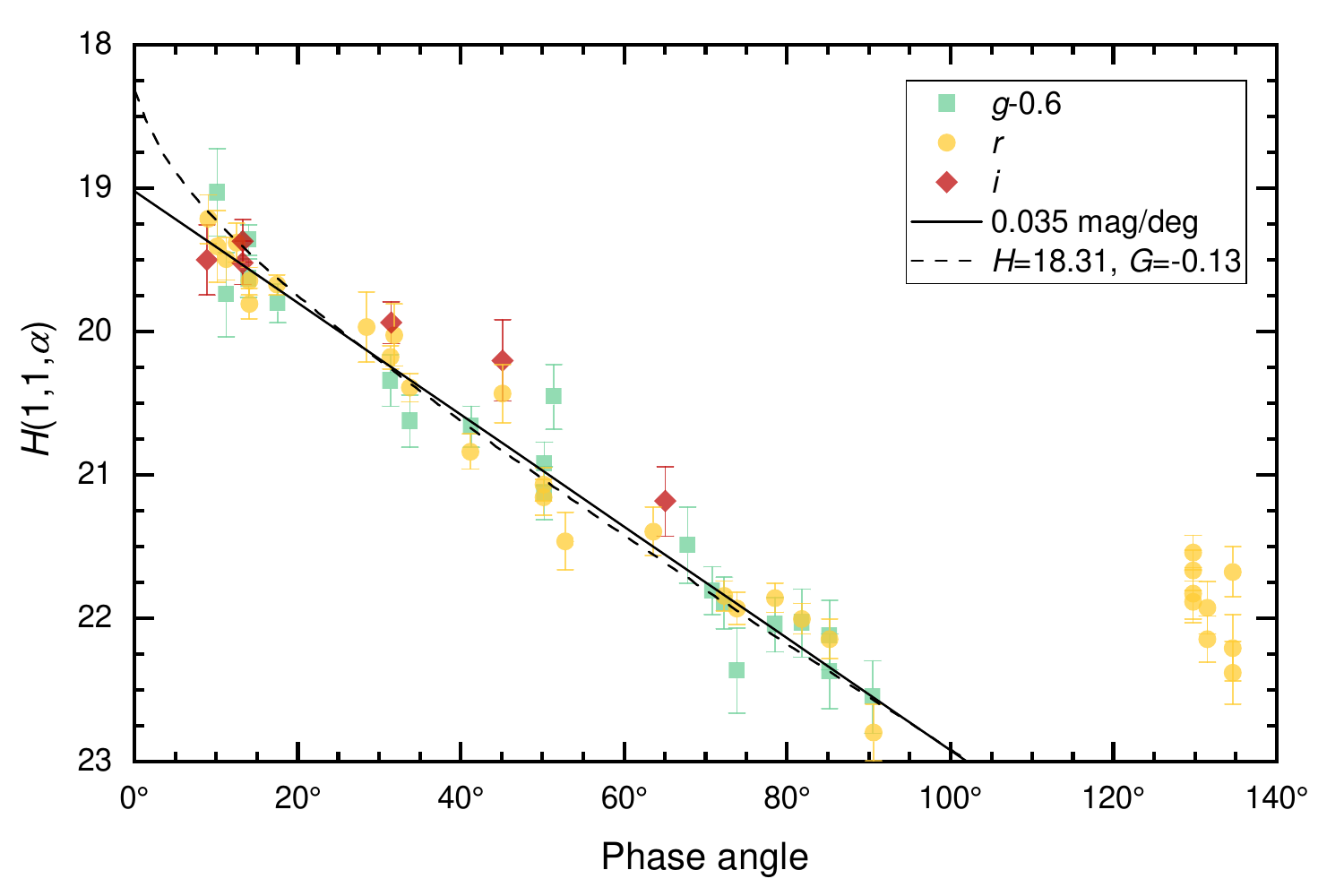}
\caption{Absolute magnitudes $H(1,1,\alpha)$ (brightness of the asteroid as if the asteroid were based at $r_\mathrm{h}=1$~au, and $\varDelta=1$~au but not corrected for the phase angle effect) derived from three-color ZTF observations as a function of phase angle. The solid line is the least-square linear fit to the $r$-band photometry with $\alpha<100^\circ$ which yields a slope of $0.035\pm0.002~\mathrm{mag/deg}$. The dashed line is an $HG$ model with $H=18.31$ and $G=-0.13$ as described in the main text. \label{fig:h11a}}
\end{center}
\end{figure*}

To test this idea, we plotted the phase function of P/2021 HS based on ZTF photometry as shown in Figure~\ref{fig:h11a}. Data from different bandpasses are aligned by a least-square fit that yields a color of $g-r=0.6\pm0.1$~mag and $r-i=0.0\pm0.1$~mag that crudely matches typical nuclei of Jupiter-family comets \citep{Jewitt2015}. We observed a linear relation within $\alpha<100^\circ$, consistent with a bare nucleus, thus showing that the comet signal in this regime was indeed dominated by the nucleus. By fitting the $r$-band data within $\alpha<100^\circ$, we obtained a phase coefficient of $0.035\pm0.002~\mathrm{mag/deg}$, compatible with typical short-period comet nuclei \citep{Kokotanekova2017}.

The data at $\alpha>100^\circ$, on the other hand, clearly deviate from the linear relation. We note that these data correspond to the observations made in early 2021 August where presence of a coma was reported. Thus, this deviation can be naturally explained by the dominance of the dust in the aperture, due to the brightness enhancement of the dust coma caused by forward scattering and/or enhanced activity at those epochs.

\begin{figure*}[h!]
\begin{center}
\includegraphics[scale=1,angle=0]{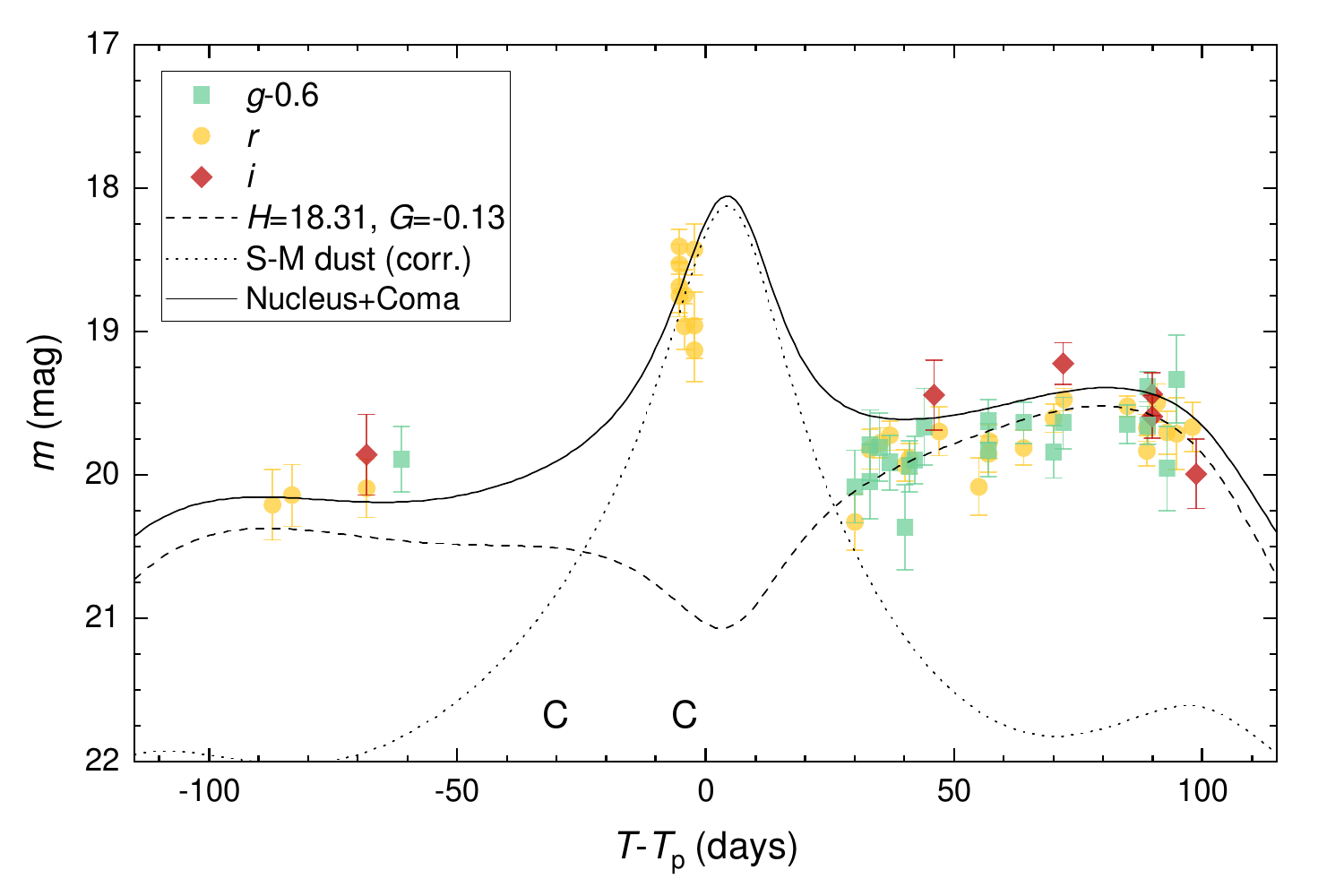}
\caption{Photometry from ZTF and LDT as well as modeled light-curves of P/2021 HS. The epochs in which a coma was detected are marked by ``C''. The dashed curve is the fitted light-curve of the nucleus assuming $H=18.1$ and $G=-0.13$, as well as a phase slope of 0.035~mag/deg as derived from Figure~\ref{fig:h11a}; the dotted curve is the Schleicher--Marcus dust function for the coma corrected for a fixed angular aperture and fitted to the $r$-band photometry of which the comet was active; the solid curve is the sum of the two light-curves. \label{fig:phot}}
\end{center}
\end{figure*}

Now that we have identified the turnover near $\alpha=100^\circ$ ($T-T_\mathrm{p}$ at -19 and +26 days, respectively), we modeled the light-curve by combining two different models: an {\it HG} model used for asteroids and bare cometary nuclei \citep[where $H$ is the absolute magnitude and $G$ is slope parameter used to describe the phase curve, cf.][]{Bowell1989}, and the Schleicher--Marcus dust model for the coma. For the $HG$ model, we assumed $G=-0.13$ following the value of 67P/Churyumov--Gerasimenko measured by {\it Rosetta} \citep{Fornasier2015} and derived $H=18.31\pm0.04$~mag, translating to a nucleus diameter of $1.45\pm0.03$~km, assuming a geometric albedo of 0.04. (Alternatively, we have $H_\mathrm{lin}=19.02\pm0.08$~mag if we ignore the opposition effect and fit the light-curve using a linear function.) If we use $H=19.02$ derived from the linear phase function model \citep[in consistent with][]{Kokotanekova2017} and consider the size measured by {\it NEOWISE} (0.6--1.1~km; see \S~\ref{sec:neowise}), we derive an albedo of 0.04--0.12, which is on the higher end of typical JFCs \citep[$\lesssim0.07$; ][]{Kokotanekova2017} and is more in line with S-type asteroids. For the Schleicher--Marcus model, a term of $-2.5\log{\varDelta}$ is added to account for the fixed angular aperture used to derive the ZTF photometry. The result, shown in Figure~\ref{fig:phot}, reveals a crude match between the periods that the coma dominates the total brightness versus that the epochs that a coma was reported, again supporting the idea that the detection of cometary activity is primarily caused by the enhancement from forward scattering of dust rather than intermittent ejection.

\begin{figure*}
\begin{center}
\includegraphics[scale=1,angle=0]{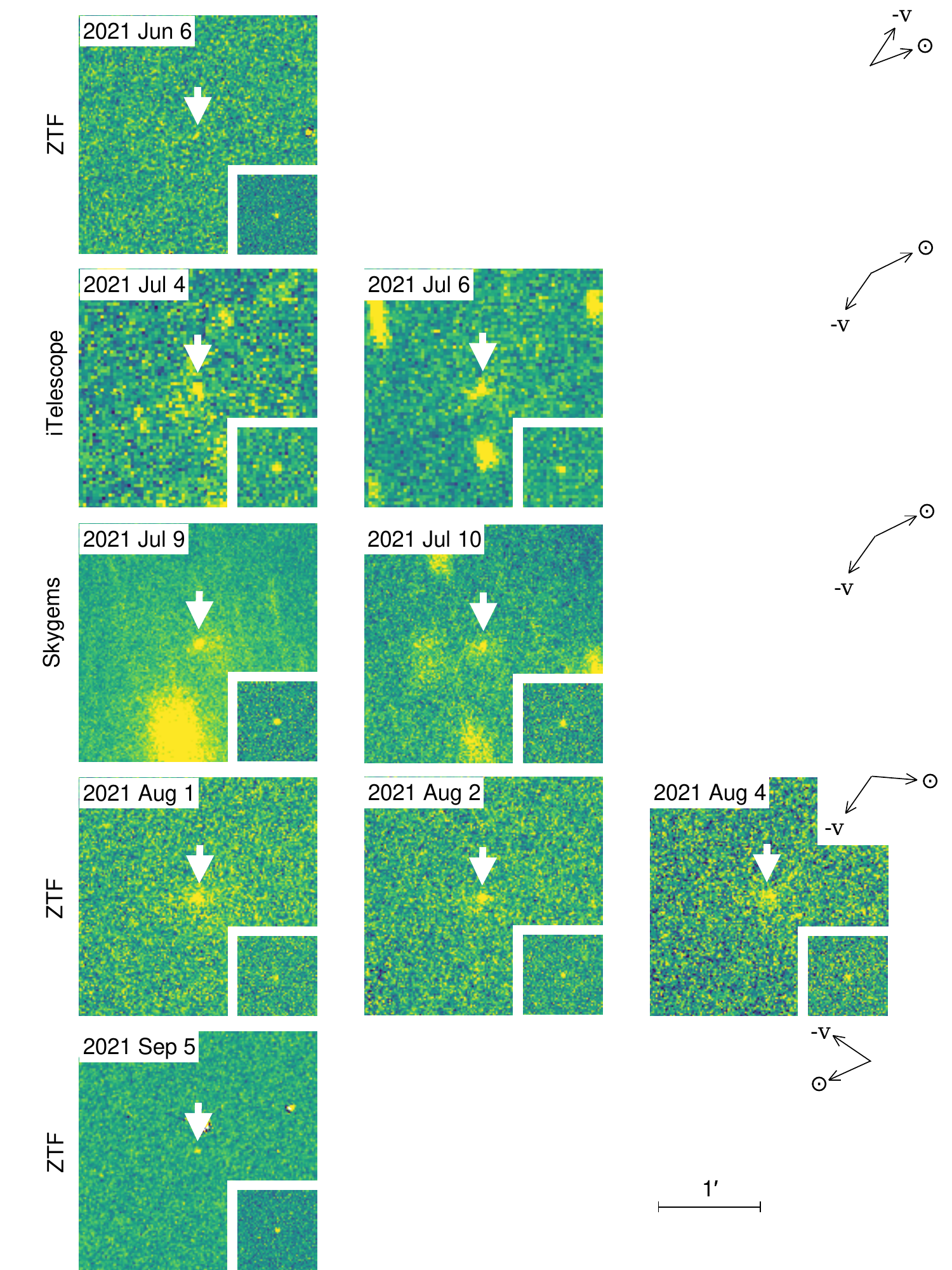}
\caption{Stacked images of P/2021 HS from 2021 June 6 to September 5, showing the comet throughout its active phase, complete with images taken immediately before and after the active phase. Point-source function (PSF) reference, taken from a nearby background star of similar brightness, is given in the lower right of each thumbnail. The Sunward ($\odot$) and anti-orbital velocity ($-v$) vectors are indicated with arrows.\label{fig:mosaic}}
\end{center}
\end{figure*}

\begin{figure*}[h!]
\begin{center}
\includegraphics[scale=1,angle=0]{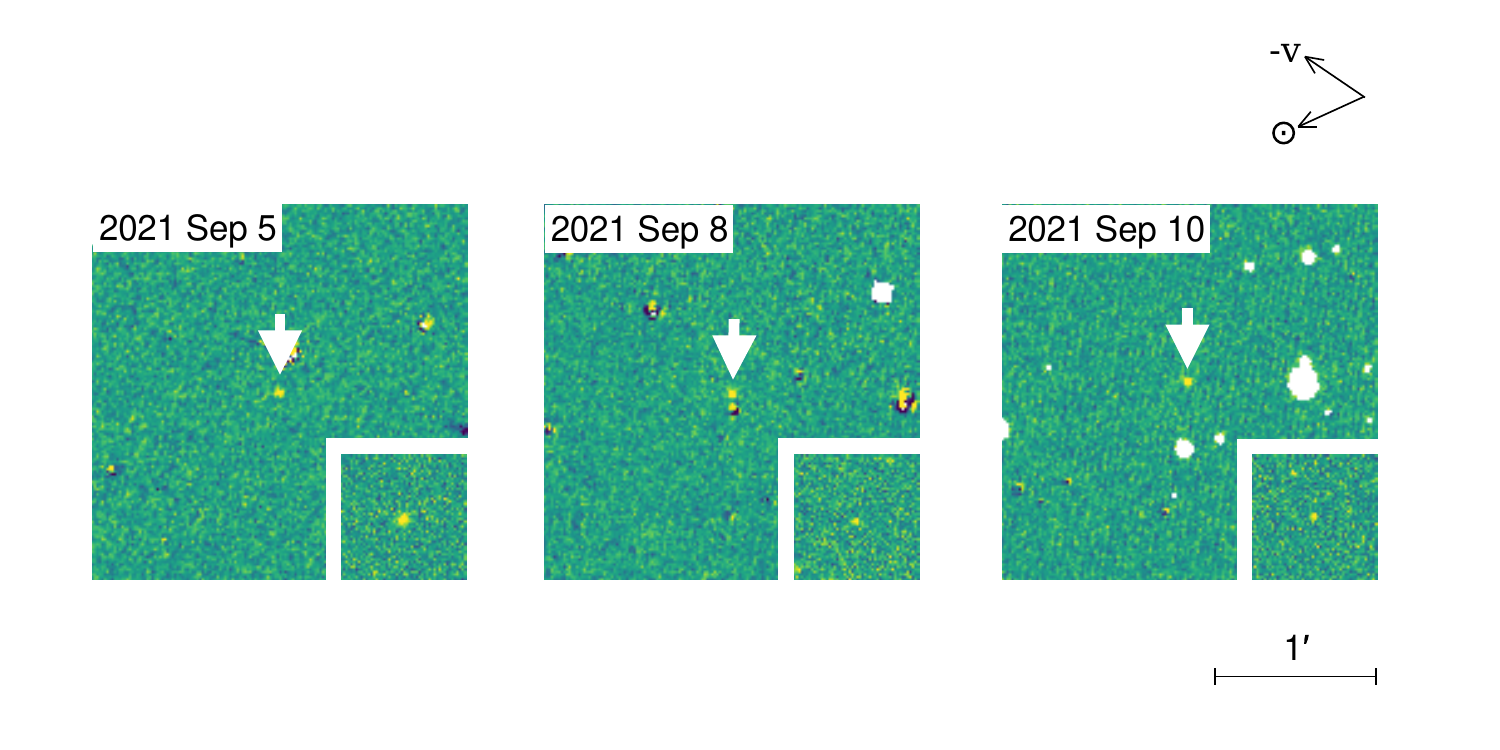}
\caption{Combined and $2\times$ binned ZTF $g+r$ star-subtracted images of P/2021 HS on 2021 September 5, 8 and 10. PSF references, taken from a nearby background star of similar brightness, are given in the lower right of each thumbnail. The sunward ($\odot$) and anti-orbital velocity ($-v$) vectors are indicated with arrows. \label{fig:ztf-binned}}
\end{center}
\end{figure*}

We note that the pre-perihelion iTelescope and Skygems images, taken at $\alpha=82^\circ$ to $91^\circ$, showed the coma, while post-perihelion ZTF observations at similar phase angles did not (Figure~\ref{fig:mosaic}). Does this indicate pre-/post-perihelion asymmetry in the activity of P/2021 HS? There are several factors to be considered. The telescopes used by iTelescope and Skygems have smaller apertures compared to ZTF (0.5~m versus 1.2~m), but the longer total integration time used by the two telescopes have largely cancelled out this difference. However, iTelescope and Skygems employed ultra-broadband $L$ filters (compared to ZTF's broadband $g$ and $r$) and have slightly larger pixel sizes (1\farcs6, 1\farcs1 vs. ZTF's 1\farcs0), both of which make them slightly more sensitive to extended structures such as a coma. As a cursory test, we combined ZTF's $g$- and $r$-band images from each of 2021 Sep 5, 8 and 10 ($\alpha=91^\circ, 85^\circ, 82^\circ$) and binned them by a factor of 2, aiming to match the sensitivity of iTelescope and Skygem. Still, no definitive coma can be seen (Figure~\ref{fig:ztf-binned}). This seems to support the idea that the comet was more active before the perihelion, and is echoed by the fact that the photometry around these dates ($T-T_\mathrm{p}\approx30$~days) are below best-fit light-curves, as shown in Figure~\ref{fig:phot}.

\begin{figure*}[h!]
\begin{center}
\includegraphics[scale=1,angle=0]{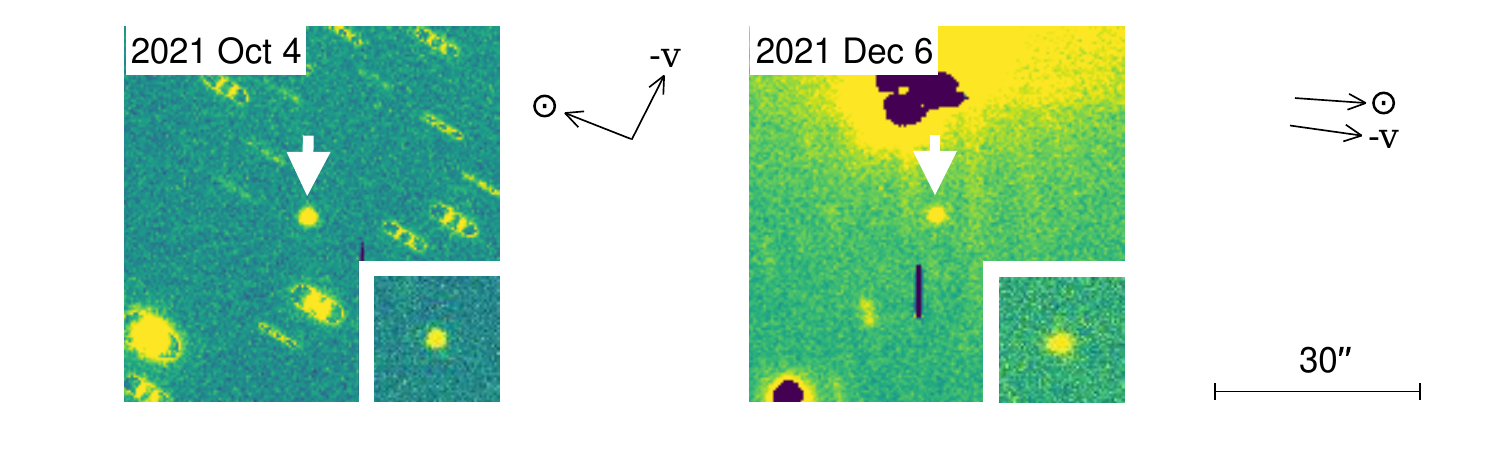}
\caption{Same as Fig.~\ref{fig:ztf-binned}, but for stacked LDT images of P/2021 HS from 2021 October 4 and December 6.  The vertical purple bars in the images are due to a detector artifact. \label{fig:ldt}}
\end{center}
\end{figure*}

We also examined the stacked LDT images for coma, but did not find one (Figure~\ref{fig:ldt}). Photometry under a $4''$-radius aperture yielded $r=19.78\pm0.03$ and $r=21.30\pm0.05$ for 2021 October 4 and December 6, respectively. This agrees with a nucleus with $H=18.31$, $G=-0.13$, and a phase coefficient of $0.035~\mathrm{mag/deg}$, within uncertainty (such a nucleus would have $r=19.69\pm0.09$ and $r=21.22\pm0.09$ on these two dates), meaning that no statistically significant coma is detected. On the other hand, the Schleicher--Marcus model suggests that the coma within a 4''-radius aperture would be $\sim2$~mag fainter than the nucleus, meaning that the combining uncertainties of the coma and nucleus photometry needs to be smaller than $5\%$ in order to detect the coma at $3\sigma$ level. Thus, it is not surprising that the coma is not detected: deeper images and better nucleus brightness estimates would have been needed.

\subsection{Coma Morphology}

\begin{figure*}
\begin{center}
\includegraphics[scale=1,angle=0]{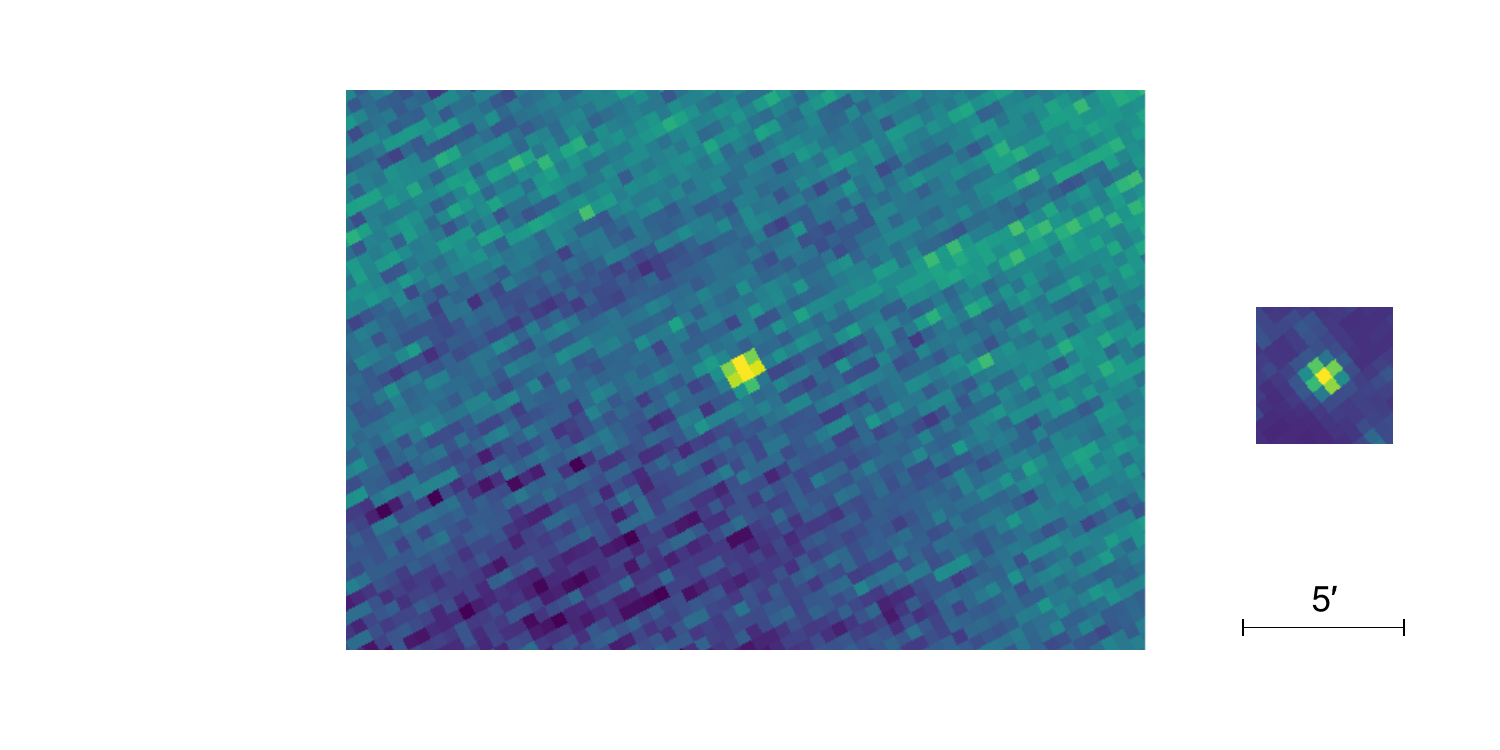}
\caption{Left: Stacked image of P/2021 HS generated using 3446 star-subtracted {\it TESS} images obtained during 2021 April 2--28, with a total integration time of 574~hours; Right: A background star as PSF reference. The orientation of the trail rotated $11^\circ$ during this period, and thus this image is only useful for the search of coma. No definitive coma can be seen.\label{fig:tess-deep}}
\end{center}
\end{figure*}

\begin{figure*}
\begin{center}
\includegraphics[scale=0.8,angle=0]{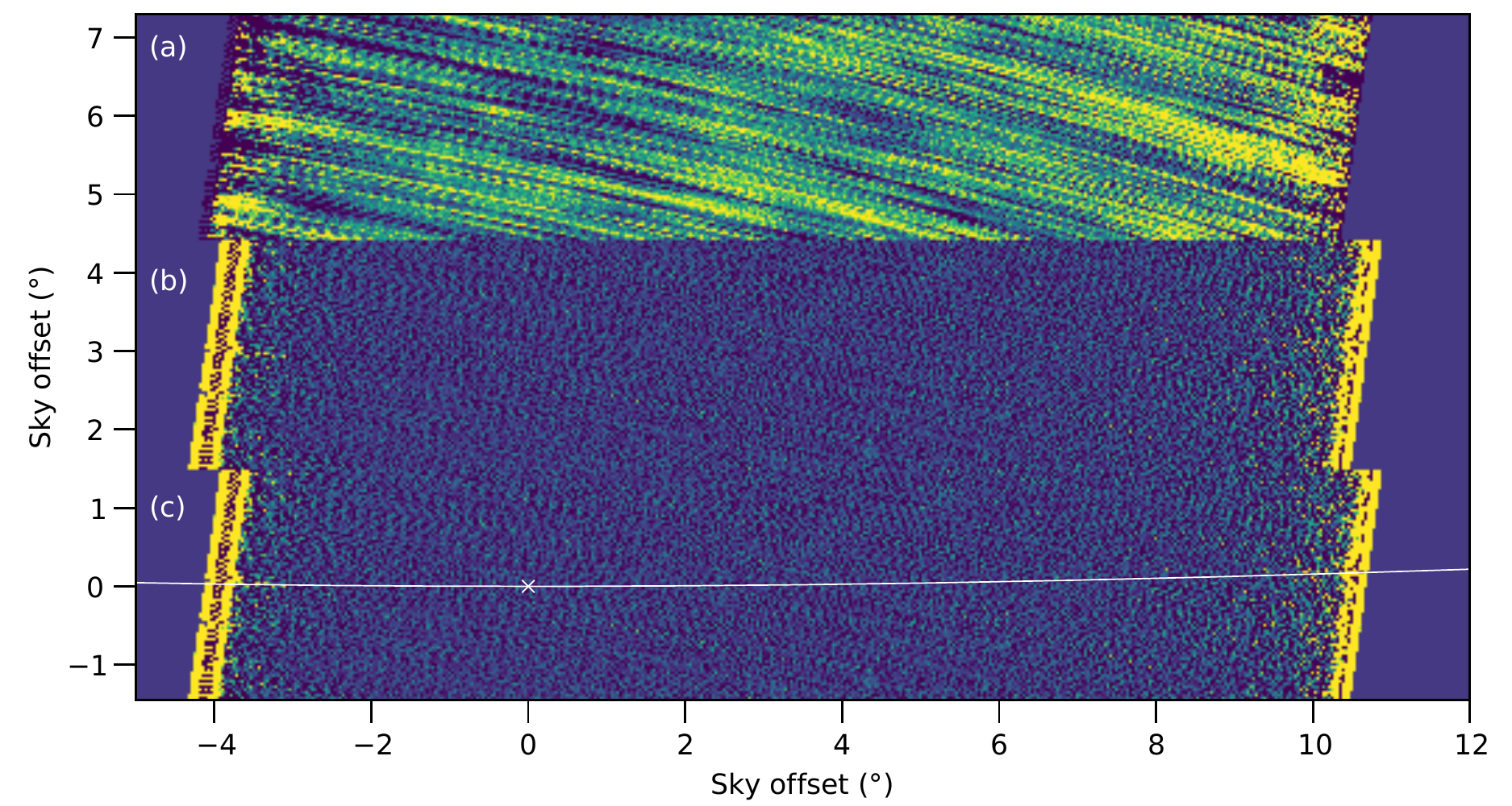}
\caption{Stacked image of P/2021 HS generated using 720 {\it TESS} images obtained during 2021 April 22--26. Compared to Figure~\ref{fig:tess-deep}, the time range of this image is carefully chosen so that the trail rotation is negligible ($\sim1^\circ$). (a) The original median stacked imaged, aligned in the rest frame of the comet. (b) The star-trail cleaned image.  (c) Same as (b), but with the position of the comet marked with an ``x'' and the projected orbit indicated with a solid line. No evidence for a dust trail is seen in our data and the comet was undetected. The non-detection of the comet as opposed to Figure~\ref{fig:tess-deep} is because we only use a small fraction of the images to avoid trail smearing due to the change of observing geometry. \label{fig:tess}}
\end{center}
\end{figure*}

Figure~\ref{fig:mosaic} shows a mosaic of stacked images from the nights when a coma was reported. We also visually inspected stacked images from other nights, as well as the deep {\it TESS} stacked image from 2021 April (Figures~\ref{fig:tess-deep}, \ref{fig:tess}), but did not find additional epochs of which the comet was active. The coma was near the detection limit of the iTelescope images on 2021 July 4 and 6, owing to the shorter integration time, but was apparent on the Skygems and ZTF images taken thereafter (though the signal-to-noise ratio was still quite modest). In all these images, the weak coma encompassed no more than a few FWHM units, or on the order of $\sim10''$. The low signal-to-noise ratio as well as the modest apparent size makes it difficult to resolve any sub-structures such as a tail or a jet. In particular, there is no clear elongation along the anti-solar or the velocity direction (marked by $\odot$ and $-v$ on the figure) that resembles a tail.

To explore the physics of the observed coma, we employed a dust dynamics model to simulate the coma morphology. In \S~\ref{sec:lc}, we have shown that the detection of comet activity is due to the brightness enhancement caused by forward scattering effect, but the nature of the underlying activity is not clear. Therefore, we consider these following scenarios: (1) sublimation-driven steady-state activity that is typical among ``normal'' comets; (2) impulsive ejection that resembles an outburst caused by rapid sublimation of volatiles; and (3) low-speed (with a terminal speed on the order of 0.2~m/s) isotropic impulsive ejection, resembling activity caused by global non-sublimatory processes such as spin-up disruptions. [Impact of an external object is another possibility, but we consider this unlikely because (a) collisions between near-Earth objects are unlikely \citep{Bottke1993}\footnote{\citet{Marzari2011} found that spin-up disruption dominates over impact disruption for km-class main-belt asteroids, while \citet{Bottke1993} showed that collisions between near-Earth objects are an order of magnitude less likely than collisions between main-belt asteroids. Assuming that the rate of spin-up disruption does not decrease closer to the Sun, we conclude that impact disruption remains an unlikely scenario for near-Earth object disruptions.}; and (b) impact disruption usually generates asymmetric ejecta aligned with the impact direction, which is not supported by the observations.] As these models involve many free and poorly-constrained quantities, we simplistically assume an isotropic ejection with a dust size range of [$10^{-4}$, $10^{-1}$]~m and that the sizes follow a power-law distribution of $\mathrm{d}n/\mathrm{d}a\propto a_\mathrm{d}^{-3.6}$, typical among comets \citep{Fulle2004}.

To simulate the sublimation-driven activity, we define a \citet{Whipple1951}-like model as follows:

\begin{equation}
    v_\mathrm{t} = \left( \frac{1}{\rho_\mathrm{d} a_\mathrm{d}} \right)^{1/2}
\end{equation}

\noindent where $v_\mathrm{t}$ is the terminal speed of the dust in m~s$^{-1}$, $\rho_\mathrm{d}=2000~\mathrm{kg~m^{-3}}$ and $a_\mathrm{d}$ is the bulk density and the diameter of the dust in m, respectively. For sustained activity, we assume that the activity starts at 3~au which is the distance that water ice, the most common species on comets, starts sublimating effectively. For impulsive activity, we assume an ejection epoch of 2021 June 15 ($T-T_\mathrm{p}=-51$~days), about mid-way between our last inactive detection and the first detection of activity. For the low-speed ejection, we assume a uniform ejection speed of 1~m/s as suggested by previous works \citep[cf.][Figure~18]{Jewitt2015}. The motion of the dust particles are then simulated with the consideration of solar radiation pressure.

\begin{figure*}
\begin{center}
\includegraphics[scale=1,angle=0]{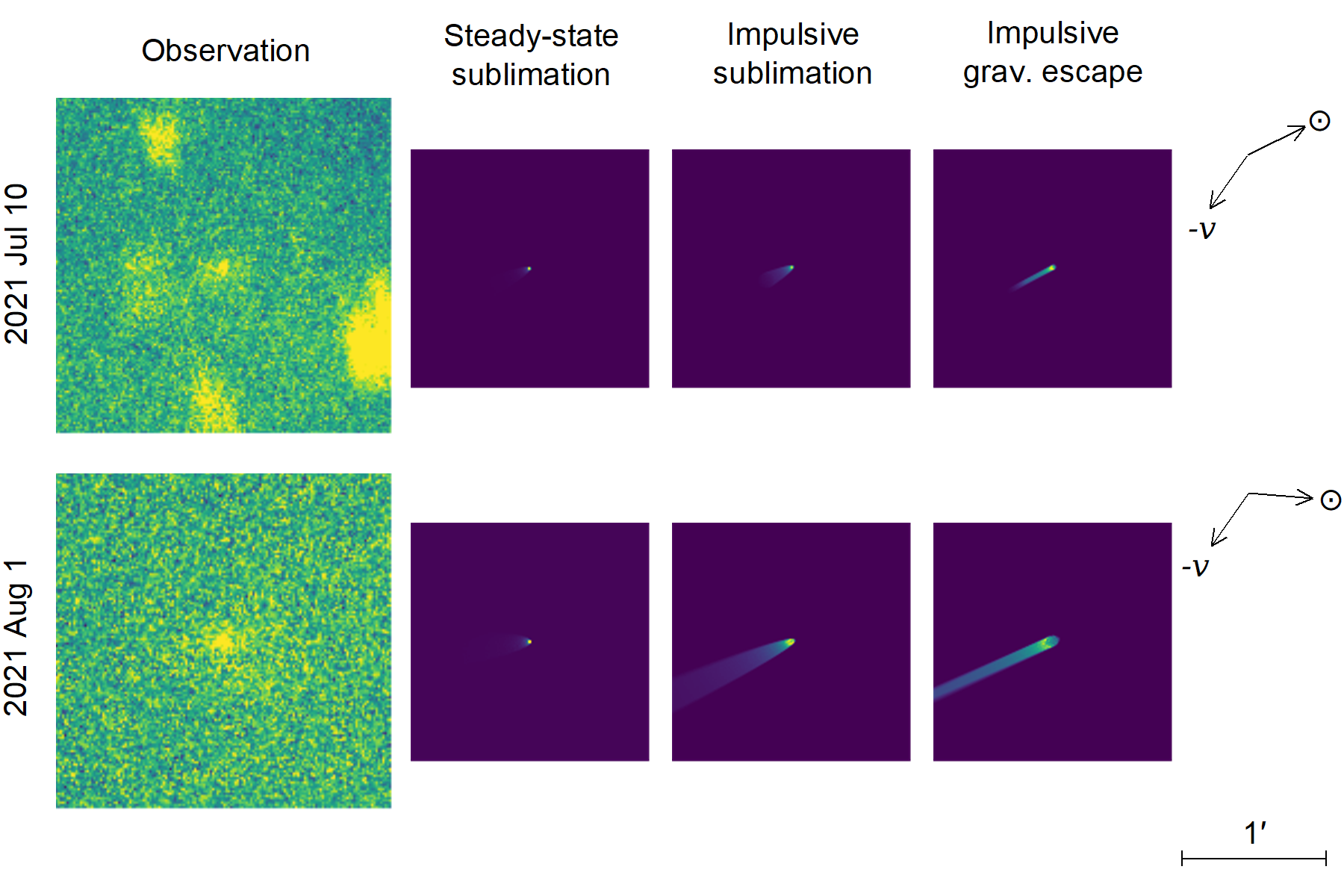}
\caption{Observed (from Fig.~\ref{fig:mosaic}) and simulated images of P/2021 HS on 2021 July 10 and August 1 under different activity scenarios (see main text). Images on each row have the same photometric scale. \label{fig:sim}}
\end{center}
\end{figure*}

Figure~\ref{fig:sim} shows the result of the simulations. The steady-state model shows a comet with a compact coma and tail, whilst the two impulsive models show a coma that becomes more diffuse over time and a longer tail. This is because the comet flux in the steady-state model is dominated by freshly-generated dust concentrated near the nucleus, while the one-off ejecta in the impulsive models are slowly blown away from the nucleus by solar radiation. The observation, while limited in spatial resolution and relatively noisy, is clearly more consistent with the steady-state model. We note that the parameter space of dust dynamics models is broad \citep[e.g.][]{Ishiguro2009}; certain combinations, such as a power-law break at unusually large dust size, can in theory reproduce a coma that satisfies the morphological constraint set by the (limited) observations. However, given that this would require an unusual dust size distribution not typically observed among comets, we consider such a scenario to be unlikely.

\subsection{Dust Trail}

Cometary dust trails are formed by large, old-age dust grains ejected by their parent comets long ago. Such structures have been detected in the optical and infrared wavelengths \citep{Sykes1992, Ishiguro2009} including ones from comets that are not highly active in dust emission \citep[e.g., 46P/Wirtanen;][]{Farnham2019}. The infrared NEOWISE data, though at a wavelength more suitable to detect dust trails, are too shallow for this purpose (and no trails were visible, as discussed in \S~\ref{sec:neowise}). The deep {\it TESS} image shown in Fig.~\ref{fig:tess} also does not show any trail structure, from which we estimate an upper limit in surface brightness of $\mu_\mathrm{TESS}=$28.7~mag/arcsec$^2$. Assuming that this is dominated by the scattered light from the dust, the optical depth of the dust trail can be written as

\begin{equation}
    \tau = C/\rho^2
\end{equation}

\noindent where

\begin{equation}
    C=\frac{\pi r_\mathrm{h}^2 \varDelta^2}{p_{TESS} \Phi(\alpha)} \frac{F_\mathrm{trail}}{F_\odot}
\end{equation}

\noindent is the cross section of the dust, with $p_\mathrm{TESS}=0.1$ is the geometric $\mathrm{TESS}$-band albedo of the dust, $F_\mathrm{trail}$ and $F_\odot$ are the $\mathrm{TESS}$-band flux of the trail in 1~arcsec$^2$ and the Sun, and $r_\mathrm{h}$, $\varDelta$ and $\Phi(\alpha)$ follows the definition given above; and $\rho$ is the linear distance at the comet that corresponds to $1''$. Inserting all the numbers, we obtained an upper limit of $\tau=3\times10^{-10}$. The total mass of the dust trail can then be constrained using \citet[][Equation~9]{Reach2007}. Taking a $40''$ (2 {\it TESS} pixels) width of the trail, a characteristic grain size of 1~mm, and a trail extent of $10^\circ$ as found by \citep[][Table~IV]{Sykes1992}, we obtained $M_\mathrm{trail} \lesssim 3\times10^{8}$~kg. Assuming that this mass is ejected over $\sim1$~orbit, this implies a mass loss rate lower than $\sim1~\mathrm{kg~s^{-1}}$. The optical depth and mass estimates are comparable to or smaller than the least massive cometary dust trails ever detected. We do, however, note that the trail extent varies from comet to comet, hence the mass estimate is probably only good to one order of magnitude.

\section{Discussion}

\subsection{Meteor Activity}

Noting that P/2021 HS has a minimum orbit intersection distance (MOID) of 0.042~au with the Earth, well within the range of comets that generate meteor showers \citep[$\sim 0.1$~au,][]{Jenniskens2021}, we searched for meteor activity originating from this comet. Using the approach described in \citet{Neslusan1998}, we calculated that the theoretical radiant of the meteor shower is $\alpha=224^\circ$, $\delta=-40^\circ$, with a geocentric speed of $19~\mathrm{km~s^{-1}}$ and a peak solar longitude of $70^\circ$ (approximately from the end of May to the beginning of June). A search in the IAU Meteor Data Center's catalog of known meteor showers\footnote{\url{https://www.ta3.sk/IAUC22DB/MDC2007/}, accessed 2022 April 3.} revealed no matches. Taking a typical detection limit of $\sim 10^{-4}~\mathrm{km^{-2}~hr^{-1}}$ for mm-class meteoroids, achievable by meteor surveys nowadays, and an activity duration of 1~day, we obtained an upper limit to the meteoroid stream mass, $\lesssim5\times10^7$~kg, compatible with the upper limit of the trail mass derived above.

We also performed a cursory meteoroid stream simulation using the approach described in \citet{Ye2016} assuming mm-class meteoroids. The result, tabulated in Table~\ref{tbl:met}, showed limited direct encounters with ejecta from specific years since 1900. Planetary dynamics will steer ejecta formed in 1895, 1903 and 1733 into Earth's orbit in the years of 2055, 2070, and 2090, potentially producing meteor outbursts. Assuming no change in the activity level of P/2021 HS, these meteor outbursts would likely be weak (with meteor flux comparable or below the flux of sporadic meteors). However, if the comet has been more active in the past, these meteor outbursts could become detectable.

\begin{deluxetable*}{lccl}
\tablecaption{Predicted significant meteor outbursts from P/2021 HS from 1900 to 2100.}
\label{tbl:met}
\tablecolumns{4}
\tablehead{
\colhead{Date/Time (UT)} & \colhead{Year of ejection} & \colhead{Duration} & \colhead{Note}
}

\startdata
2055 May 26 08:16 & 1895 & 1~hr & ZHR$\lesssim2$ \\
2070 May 21 12:18 & 1895, 1903 & 1~hr & ZHR$\lesssim2$ \\
2090 May 26 22:16 & 1733 & 3~hr & ZHR$\lesssim2$ \\
\enddata
\end{deluxetable*}

\subsection{Nature}

Given that the comet signal in the 2021 August ZTF images is dominated by coma flux (as suggested by Figure~\ref{fig:phot}), we estimated from Figure~\ref{fig:afrho} that a nucleus-free $A(0^\circ)f\rho \approx 0.001$~m. This is similar to the $<0.0019$~m estimated for P/2016 BA$_{14}$ measuring during its active period \citep{Li2017}, a comet that has been recognized as an extremely low-activity comet \citep{Li2017}. The measurement for P/2021 HS is the lowest value ever measured on a comet near 1~au. Assuming the same gas-to-dust ratio as 209P/LINEAR, another extremely low-activity comet with $A(0^\circ)f\rho \approx 0.01$~m \citep{Ye2014, Schleicher2016}, this converts to an active area of $\sim700~\mathrm{m^2}$ or $\sim0.02-0.06\%$ of a spherical surface taken our previous diameter estimate of 0.6--1.1~km. This translates to a single circular pit with a 15-m radius.

\begin{figure*}
\begin{center}
\includegraphics[scale=1,angle=0]{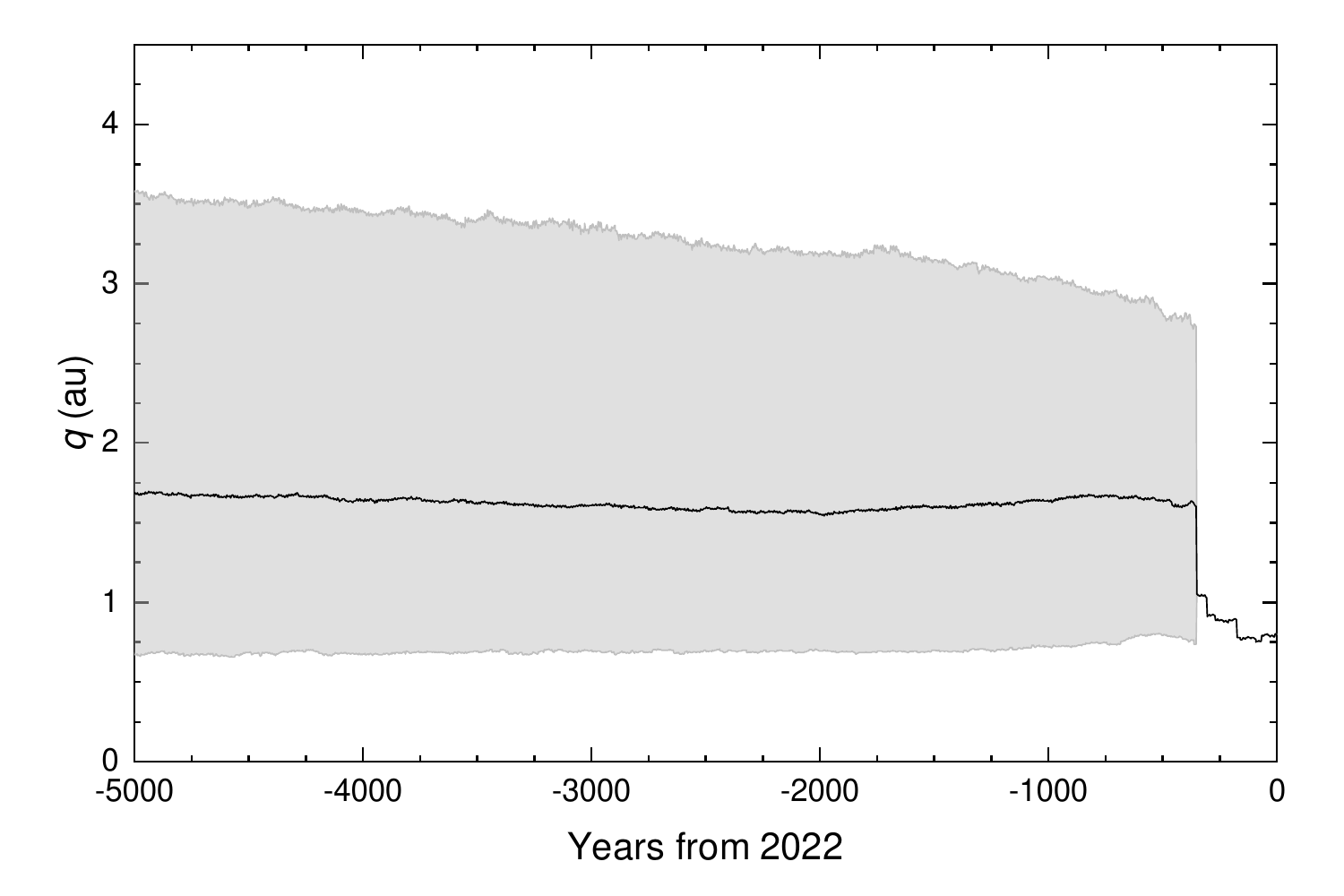}
\caption{Evolution of perihelion distance $q$ of P/2021 HS in the past 5000~yr. The shaded area are $3\sigma$ band of the clones, generated and integrated using orbit solution JPL \#39.\label{fig:dyn}}
\end{center}
\end{figure*}

Is P/2021 HS a near-death comet that has depleted all its volatiles, or did it form in a region closer to the Sun than typical comets where less volatiles were available? The chaotic nature of comet dynamics does not permit a definitive answer, but the present orbit of the comet can provide some clues. Despite the fact that P/2021 HS has only been observed over one apparition, the available orbit is good enough to confidently trace the comet back to its last planetary encounter, a $\sim0.05$~au encounter with Jupiter in 1670 January, as shown in Figure~\ref{fig:dyn}. An encounter this close erases the dynamical memory quite effectively, hence it is unlikely that further improvement on the orbit can recover the dynamical history further. However, this encounter alone shows that P/2021 HS is prone to close encounters with Jupiter, a trait that is common among ``typical'' JFCs of which are thought to have originated from the trans-Neptunian region \citep[e.g.][]{Fernandez1980}. In contrast, many other low activity comets such as 209P/LINEAR are found to be on stable orbits over the timescale of $10^4$~years and may have originated from a region closer to the Sun, such as the asteroid belt \citep{Fernandez2015, Ye2016b}. Since P/2021 HS does not appear to follow this paradigm, we conclude that the comet is most likely from the trans-Neptunian region and has depleted most of its near-surface volatiles, likely due to many passages through the inner Solar System.

\subsection{The Challenge of Detecting Low-Activity Comets}

The ever-improving capabilities of sky surveys and other facilities have turned up many objects that lie between the traditional definition of asteroids and comets, such as active asteroids (objects in asteroidal orbits that show comet-like features) and asteroids in cometary orbits (ACOs). However, the current survey network is mostly comprised by telescopes $\lesssim2$~m in diameter that employ short exposures ($\lesssim1$~min), which are insufficient to detect extremely low-active comets. The cometary nature of P/2016 BA$_{14}$, for example, was first recognized with a 5~min exposure on the 4.3-m LDT \citep{Knight2016}. It would be difficult to realize the cometary nature of P/2021 HS had it not been observed at high phase angles. 

How many ACOs are unrecognized comets with extremely low activity? Here we consider the perspective of detecting the forward-scattering-enhanced dust coma of these comets. The Schleicher--Marcus model predicts that the forward scattering of dust coma becomes significant when $\alpha \gtrsim 120^\circ$. For Earth-based observers, this geometry is restricted to objects with $q\lesssim1$~au. To figure out how many ACOs have been observed at such high phase angle, we identified 927 ACOs (defined as asteroids with $T_\mathrm{J}<3$) with $q<1$~au, and searched the MPC database for observations made when $\alpha>120^\circ$. Observations of 12 objects, observed in 17 different nights, were found, of which none has been reported to show a coma or a tail. We also note that P/2021 HS is, to our knowledge, the only example of a comet identified by dust coma enhanced by significant forward scattering effect. Assuming random sampling and that P/2021 HS is the sole ``positive'' detection of such extremely-low-activity comets, we have a positive detection rate of $\sim8\%$. This is in line with the constraint from meteor observations that at least $10\%$ of the near-Earth ACOs are dormant comets \citep{Ye2016a}. On the other hand, as much as $\sim50\%$ of ACOs have physical properties compatible with cometary nuclei \citep{Mommert2015, Mommert2020}. These two piece of evidences imply that the majority of true comets in the ACO population are in dormancy or very nearly so. However, we should immediately note that this only represents $12/927\approx1\%$ of the selected ACOs, and thus any conclusion drawn from this very small sample is weak.

Another way to detect low-activity comets is through their gas emission. This does not have the benefit of the forward scattering enhancement, but it is easier to isolate emission lines or bands using certain techniques (e.g., spectroscopy or narrow-band filters centered on the emission lines or bands) which is not possible for continuum features such as dust emissions. An example is the cometary activity of (3552) Don Quixote which was first detected through its CO and/or CO$_2$ emission \citep{Mommert2014}. Subsequent deep optical imaging of Don Quixote revealed its very faint dust coma \citep{Mommert2020b}, supplied by an active area of $6500~\mathrm{m^2}$. A reconnaissance survey of 75 selected ACOs using the same approach used to discovery the activity of Don Quixote did not reveal additional comets \citep{Mommert2020}. Another attempt in search of CN emission from active asteroid (3200) Phaethon was conducted by \citet{Ye2021}, taking advantage of an extremely close approach of the asteroid to the Earth (0.07~au). Despite negative results, they were able to place an upper limit of an active area of $100-4000~\mathrm{m^2}$ depending on the assumptions, on par with the measured active area of P/2021 HS. Therefore, deep narrowband imaging is a useful technique to look for low-activity comets, and can match the sensitivity of high phase angle observation under certain circumstances.

\section{Conclusions}

Jupiter-family comet P/2021 HS (PANSTARRS) is, in terms of dust production rate at perihelion, among the comets with lowest level of activity ever observed. Optical phase curve and thermal observations show a nucleus with diameter between 0.6 and 1.1~km and an albedo between 0.04 and 0.12. The extremely thin coma was only detected thanks to the strong forward scattering enhancement during a $\alpha=139^\circ$ conjunction in its 2021 apparition. The measured coma flux translates to an active area of only $700~\mathrm{m^2}$ and is among the lowest values ever measured on a comet, an order of magnitude smaller than several notable low-activity comets such as 209P/LINEAR and (3552) Don Quixote\footnote{\citet{Li2017} estimated the active fraction of P/2016 BA$_{14}$ to be $\sim0.01\%$, or $70~\mathrm{m^2}$ in terms of active area, and hence P/2016 BA$_{14}$ is technically the comet with smallest active area ever measured. However, measurements of both P/2016 BA$_{14}$ and P/2021 HS were made near the detection limit and thus are associated with appreciable uncertainties, and were derived using different approaches: narrowband imaging for the CN gas was used for the case of P/2016 BA$_{14}$, and broadband imaging for the dust coma enhanced by forward scattering for the case of P/2021 HS. Hence, it is difficult to compare one with another.}. A search for dust trails and meteor activity turned up empty, implying an upper limit in mass of the dust trail to be $\lesssim10^8$~kg, in line with the hypothesis that the comet has been nearly inactive for some time.

Dynamical analysis shows that P/2021 HS has occasional encounters with Jupiter, and the most recent one occurred in 1670 at a distance of 0.05~au. This is in line with the idea that the comet is a typical Jupiter-family comet with an origin in the trans-Neptunian region. The extremely low level of activity is likely the result of extreme depletion of near-surface volatiles caused by excessive passages in the inner Solar System. Though we caution that there are several examples of ``resurrection'' of seemingly low activity comets, such as 252P/LINEAR, 297P/Beshore and 332P/Ikeya-Murakami \citep{Li2017,Ye2017}, and we cannot exclude the possibility that P/2021 HS will join the line some time in future.

Comets with activity this low presently are difficult to detect with conventional broadband surveys, even with deep exposures. Surveys that focus on twilight directions \citep[e.g.,][]{Seaman2018, Ye2020} are prone to the detection of comets at high phase angles and may help identify comets with extremely low activity. Alternatively, deep narrowband imaging can isolate the signal of emission lines/bands and can rival the sensitivity of high phase angle searches when conditions are right.

\bigbreak

We thank Matthew Knight, Zhong-Yi Lin, David Kaplan for discussions and suggestions that improve this paper. We also thank two anonymous referees for their careful reviews and valuable comments. We are grateful to Ben Shafransky, Cecilia Siqueiros, and Ishara Nisley for their operational support with the Lowell Discovery Telescope, as well as Lukas Demetz, Hidetaka Sato, Patricia Arevalo, and Alfredo Zenteno for providing their data for our analysis. QY, MSPK and TF acknowledge support from NASA Solar System Workings award 80NSSC21K0156.

Based on observations obtained with the Samuel Oschin Telescope 48-inch and the 60-inch Telescope at the Palomar Observatory as part of the Zwicky Transient Facility project. ZTF is supported by the National Science Foundation under Grant No. AST-2034437 and a collaboration including Caltech, IPAC, the Weizmann Institute of Science, the Oskar Klein Center at Stockholm University, the University of Maryland, Deutsches Elektronen-Synchrotron and Humboldt University, the TANGO Consortium of Taiwan, the University of Wisconsin at Milwaukee, Trinity College Dublin, Lawrence Livermore National Laboratories, and IN2P3, France. Operations are conducted by COO, IPAC, and UW. 

This research used the facilities of the Canadian Astronomy Data Centre operated by the National Research Council of Canada with the support of the Canadian Space Agency.

This project used data obtained with the Dark Energy Camera (DECam), which was constructed by the Dark Energy Survey (DES) collaboration. Funding for the DES Projects has been provided by the US Department of Energy, the US National Science Foundation, the Ministry of Science and Education of Spain, the Science and Technology Facilities Council of the United Kingdom, the Higher Education Funding Council for England, the National Center for Supercomputing Applications at the University of Illinois at Urbana-Champaign, the Kavli Institute for Cosmological Physics at the University of Chicago, Center for Cosmology and Astro-Particle Physics at the Ohio State University, the Mitchell Institute for Fundamental Physics and Astronomy at Texas A\&M University, Financiadora de Estudos e Projetos, Funda\c{c}\~{a}o Carlos Chagas Filho de Amparo \`{a} Pesquisa do Estado do Rio de Janeiro, Conselho Nacional de Desenvolvimento Cient\'{i}fico e Tecnol\'{o}gico and the Minist\'{e}rio da Ci\^{e}ncia, Tecnologia e Inova\c{c}\~{a}o, the Deutsche Forschungsgemeinschaft and the Collaborating Institutions in the Dark Energy Survey.

The Collaborating Institutions are Argonne National Laboratory, the University of California at Santa Cruz, the University of Cambridge, Centro de Investigaciones En\'{e}rgeticas, Medioambientales y Tecnol\'{o}gicas–Madrid, the University of Chicago, University College London, the DES-Brazil Consortium, the University of Edinburgh, the Eidgen\"{o}ssische Technische Hochschule (ETH) Z\"{u}rich, Fermi National Accelerator Laboratory, the University of Illinois at Urbana-Champaign, the Institut de Ci\`{e}ncies de l’Espai (IEEC/CSIC), the Institut de F\'{i}sica d'Altes Energies, Lawrence Berkeley National Laboratory, the Ludwig-Maximilians Universit\"{a}t M\"{u}nchen and the associated Excellence Cluster Universe, the University of Michigan, NSF's NOIRLab, the University of Nottingham, the Ohio State University, the OzDES Membership Consortium, the University of Pennsylvania, the University of Portsmouth, SLAC National Accelerator Laboratory, Stanford University, the University of Sussex, and Texas A\&M University.

Based on observations at Cerro Tololo Inter-American Observatory, NSF's NOIRLab (NOIRLab Prop. IDs 2020A-0909 and 2021A-0149; PIs: Arevalo, Zentedo), which is managed by the Association of Universities for Research in Astronomy (AURA) under a cooperative agreement with the National Science Foundation.

The Guide Star Catalogue-II is a joint project of the Space Telescope Science Institute and the Osservatorio Astronomico di Torino. Space Telescope Science Institute is operated by the Association of Universities for Research in Astronomy, for NASA under contract NAS 5–26555. The participation of the Osservatorio Astronomico di Torino is supported by the Italian Council for Research in Astronomy. Additional support was provided by European Southern Observatory, Space Telescope European Coordinating Facility, the International GEMINI project and the European Space Agency Astrophysics Division.

Pan-STARRS is supported by the National Aeronautics and Space Administration under Grant No. 80NSSC18K0971 issued through the SSO Near Earth Object Observations Program.

This paper includes data collected with the {\it TESS} mission, obtained from the MAST data archive at the Space Telescope Science Institute (STScI). Funding for the {\it TESS} mission is provided by the NASA Explorer Program. STScI is operated by the Association of Universities for Research in Astronomy, Inc., under NASA contract NAS 5–26555.

These results made use of the Lowell Discovery Telescope (LDT) at Lowell Observatory. Lowell is a private, non-profit institution dedicated to astrophysical research and public appreciation of astronomy and operates the LDT in partnership with Boston University, the University of Maryland, the University of Toledo, Northern Arizona University and Yale University. The University of Maryland observing team consisted of Quanzhi Ye, James Bauer, Michaela Blain, Adeline Gicquel-Brodtke, Tony Farnham, Lori Feaga, Michael Kelley, and Jessica Sunshine. 

This research has made use of data and/or services provided by the International Astronomical Union's Minor Planet Center.

This research made use of Montage. It is funded by the National Science Foundation under Grant Number ACI-1440620, and was previously funded by the National Aeronautics and Space Administration's Earth Science Technology Office, Computation Technologies Project, under Cooperative Agreement Number NCC5-626 between NASA and the California Institute of Technology.

%

\vspace{5mm}
\software{{{\tt Astropy} \citep{Astropy2013}, \tt PHOTOMETRYPIPELINE} \citep{Mommert2017}, {\tt sbpy} \citep{Mommert2019}, {\tt ZChecker} \citep{Kelley2019}}
\facilities{Blanco, DCT, NEOWISE, PO:1.2m, PS1, TESS}




\bibliography{sample631}{}
\bibliographystyle{aasjournal}



\end{CJK*}
\end{document}